%% file: main.tex
\documentclass[%
preprint,
longbibliography,
amsmath,amssymb,
aps,
prb,
floatfix,
]{revtex4-2}

\usepackage[english]{babel}
\babeladjust{autoload.bcp47 = on}

\usepackage[mathlines]{lineno}

\usepackage{graphicx}
\usepackage{xcolor}
\usepackage[hidelinks]{hyperref}

\usepackage{physics}
\newcommand{\ITMO}{Department of Physics and Engineering, ITMO University, 191002, Lomonosova St. 9, St. Petersburg, Russia}

\newcommand{\Unibs}{Department of Information Engineering, University of Brescia, Via Branze 38, 25123 Brescia, Italy}

\begin{document}

\title{Mie Optical Computing}
\author{Vsevolod Kleshchenko$^1$}

\author{Vladimir Igoshin$^1$}
\author{Costantino De Angelis$^2$}
\author{Mihail Petrov $^1$}
\affiliation{$^1$\ITMO}
\affiliation{$^2$\Unibs}

\begin{abstract}

Optical computing is emerging as a promising paradigm for next-generation information processing. Diffractive optical processors rely on spatially distributed trainable degrees of freedom, leading to extended architectures. Here, we propose a compact neuromorphic optical-computing approach where the entire trainable transformation is implemented by a single Mie scatterer. By formulating computation in vector spherical harmonics basis, trainable modal couplings can be concentrated within a finite object through its T-matrix. Since available T-matrix parameters scale as the fourth power of maximal multipole order, this architecture can overcome trainable-parameter-density limitations of conventional spatially distributed diffractive processors. We demonstrate classification of phase-encoded MNIST images using scattered-field intensity. At a particle size parameter of $ka = 15$, the trained T-matrix reaches approximately 90\% test accuracy, comparable to a single-layer artificial neural network. Similar performance can be achieved using near-fields, enabling on-chip integration. We show how reciprocity, passivity, and particle symmetry constrain performance: passivity reduces the accessible operator space while improving robustness, whereas symmetry reduces the number of independent parameters. Finally, we inverse-design a non-absorbing dielectric scatterer that realizes the classification task with 84\% accuracy. These results demonstrate that nontrivial neuromorphic transformations can be encoded within the multipolar response of a single compact scatterer.

\end{abstract}

\maketitle

\section{Introduction}
Free-space optical neural networks have been rapidly developing over the last several years as one of the most promising approaches for analog optical computing as an alternative to  electronic architectures~\cite{momeni2025training}, although whether this translates into a genuine computational advantage over digital electronics -- and what role metamaterial and metasurface platforms specifically can play in achieving it -- remains an open and actively debated question~\cite{li2024exploring}. Such optical systems offer information processing at the speed of light with intrinsically low latency, massive parallelism, and potentially energy saving~\cite{Silva2014,hu_diffractive_2024,shastri2021photonics}. While the basic principles of optical computing have been established for several decades~\cite{goodman_introduction_1988,wagner_multilayer_1987} 
the rapid progress in the diffractive optics and metaoptics technologies has made possible  wider spread of optical neural networks and formation  of the modern architecture of diffractive optical neural networks (DNN) based on cascaded passive or active diffractive layers~\cite{lin_all-optical_2018}.  The spatial diffraction and scattering channels, with the latter potentially outnumbering the open diffraction channels~\cite{hoang2026mie}, play the role of communication channels and define the connectivity between the neurons~\cite{miller_communicating_2000,Silva2014}, while the trainable parameters are encoded in the local amplitude and phase response of the diffractive layers. This architecture has been successfully applied to a broad range of computational tasks, including image classification~\cite{lin_all-optical_2018,mengu_analysis_2020,sun_review_2023}, object recognition~\cite{zhou_large-scale_2021}, turbid media imaging~\cite{luo_computational_2022}, polarization ~\cite{li_universal_2023} and spectral ~\cite{hu_diffractive_2024}  processing. Additional momentum to the field was given by the development of metaoptics and metasurfaces, which enabled subwavelength control of optical states for image processing improving the performance of diffractive optical neural networks  \cite{hu_metaoptics_2026,shu_all-optical_2026}  and making them more compact.     

Despite these advances, most existing optical neural-network architectures remain fundamentally distributed systems in which increasing computational complexity requires additional diffractive layers, extending free-space propagation distances, cascaded Fourier components, or multilayer metasurface stacks~\cite{hu_diffractive_2024}. More generally, diffraction and the overlapping nonlocality of an optical transformation impose a minimum transverse size and, for sufficiently nonlocal operations, a minimum system thickness~\cite{miller_why_2023,li_spatial_2025}. On the way of downscaling the physical footprint of optical computation unit a few approaches have been suggested such as using a single scatterer for integral transformation \cite{goh_nonlocal_2022},  multiple linear \cite{luo_volumetric_nodate} and nonlinear \cite{wang_large-scale_2024} scattering in disordered microstructures, as well as scattering based nonlinear computing~\cite{wanjura2024fully} and programmable multimode propagation~\cite{onodera2026arbitrary}. However, the question of how to implement complex trainable transformations in a single compact optical element while keeping the required connectivity remains largely unexplored.


\begin{figure}[t]
    \centering
    \includegraphics[width=\columnwidth,height=0.48\textheight,keepaspectratio]{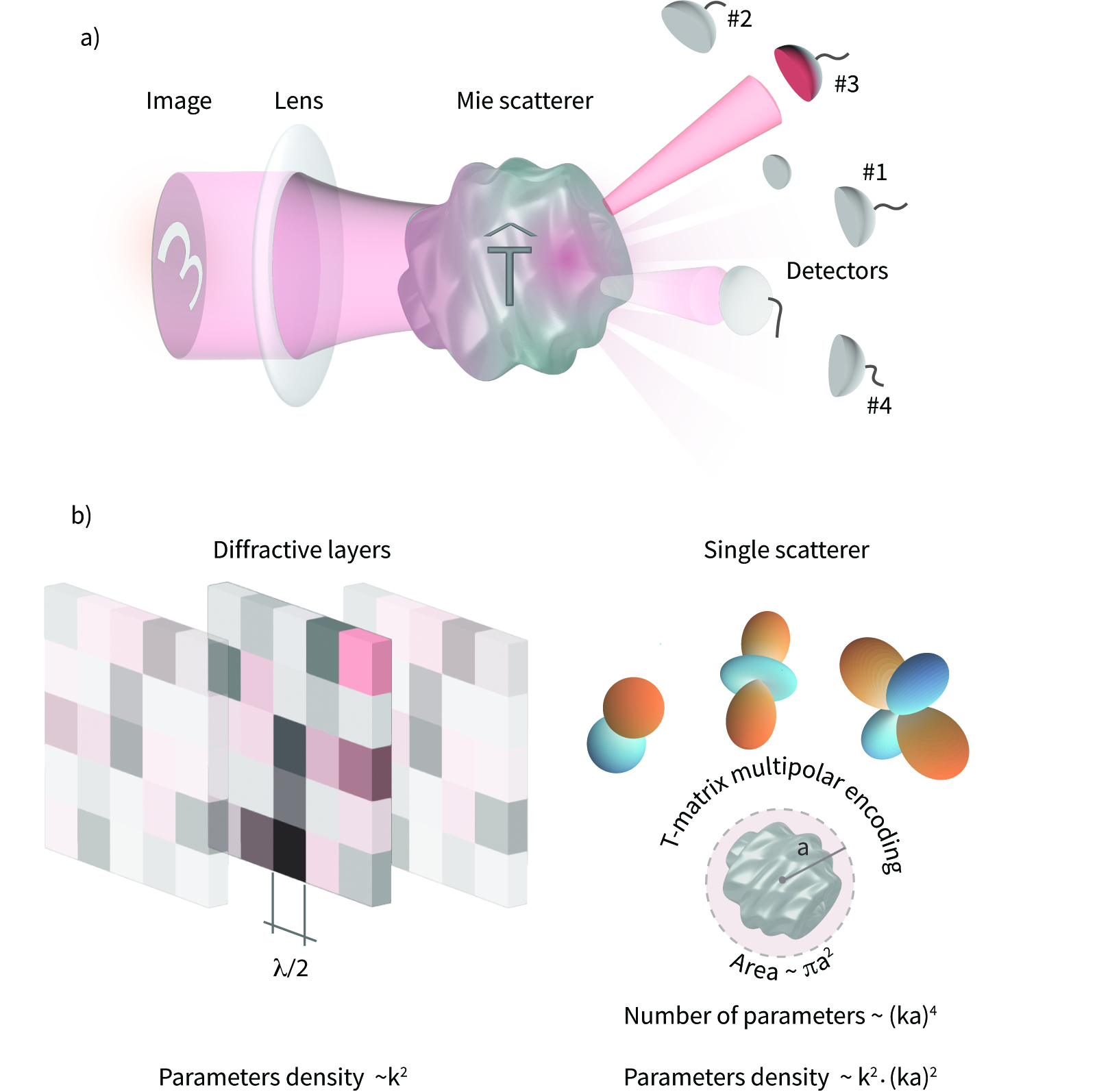}
    \caption[Concept and parameter density of optical computing with a single Mie scatterer]{
        The concept of optical computing with a single Mie scatterer with a  trainable multipolar transformation matrix for optical inference.
        (a) A phase-encoded image is focused onto a Mie scatterer whose T-matrix couples incident and scattered multipolar channels. Spatially separated detectors collect the scattered power and assign the predicted class.
        (b) Comparison of trainable-parameter densities. A conventional diffractive layer provides locally tunable pixels with a density of order $1/\lambda^2\sim k^{2}$. A scatterer of size $a$ supports approximately $(ka)^2$ multipolar channels and $(ka)^4$ pairwise T-matrix couplings, corresponding to a parameter density increase of the order of~$(ka)^2$.
        }
    \label{fig:figure0}
\end{figure}

In this work, we explore an alternative approach to  optical neural computation in which the entire trainable transformation is implemented by a single scatterer compared to a wavelength and, thus, operating in the Mie regime \cite{mie_beitrage_1908}. Multipolar nanophotonics enables control of the optical response through the excitation and interference of electromagnetic multipoles~\cite{liu2017multipolar}. Instead of distributing computation across multiple diffractive layers, we represent the optical processor by a trainable electromagnetic T-matrix \cite{mishchenko_t-matrix_1996} acting in the basis of vector spherical multipoles (VSH).  The basic principle of this approach are illustrated in Fig.~\ref{fig:figure0} (a) , where the input and output optical signals are represented as structured far-field illumination and scattered radiation, respectively, while the trainable transformation is implemented  a single compact scatterer within the T-matrix formalism. The proposed approach encodes the computational operation into the modal scattering response of a compact object, where the channel mixing in the multipolar basis can occur within a single scatterer implied by the symmetry and geometry of the structure \cite{Gladyshev2020,poleva_multipolar_2023}.

The  recent work has highlighted that the performance and scalability of optical computing systems are fundamentally constrained by the physics of wave propagation and scattering. In particular, limits associated with spatial complexity and communication-channel overlap \cite{miller_communicating_2000,miller_why_2023}, as well as optical connectivity \cite{li_spatial_2025}, constrain the size of a system implementing a given optical transformation or inference task. In this prospect, the single scatterer has much higher density of trainable parameters compared to the diffractive optical neural networks. 
Indeed, the DNN is characterized by a single ``pixel" of typical size as small as $
\sim\lambda/2$, which provides the density of trainable parameters $\sim 1/\lambda^2$ per unit area (see Fig.~\ref{fig:figure0} (b)). 
In contrast, the number of free parameters of a finite scatterer in the VSH basis depends on the number of free parameters on the T-matrix which scale as $\sim n_{\max}^4$, where $n_{\max}$ is the  multipole truncation number. The $n_{\max}\propto ka$ \cite{wiscombe_improved_1980, zhang_vector_2022},  where $a$ is the typical size of the scatterer, and $k=2\pi/\lambda$ is the wavenumber (see Fig.~\ref{fig:figure0} (b)). 
Physically, $n_{\max}^2\sim(ka)^2$ determines the number of accessible multipolar channels, whereas the $n_{\max}^4$ scaling of the T-matrix originates from pairwise couplings between the input and output channel spaces ~\cite{miller_tunnelling_2025, overvig_nonlocal_2025}.
Thus, the ratio between the number of free parameters in layered DNN and single scatterer scale as $\sim (ka)^2$ (see more details in Methods Sec.~\ref{sec:dop}) that gives already two order of increase for particle comparable to the wavelength $a\sim \lambda/2$. That opens a promising way towards new DNN architectures based on compact scatterers which can implement nontrivial neuromorphic transformations with a high density of trainable parameters.  

In the following, we consider the optical classification task of phase-encoded MNIST images \cite{lecun1998gradient} using the detected scattered-field intensity in the far- and near-fields. We demonstrate that a single scatterer can be trained to perform nontrivial classification tasks, achieving test accuracy comparable to a single-layer artificial neural network. We further investigate how physical constraints, such as reciprocity, passivity, and particle symmetry, affect the achievable performance of the T-matrix-based optical classifier. Finally, we present an inverse-designed non-absorbing dielectric scatterer that realizes the classification task with high accuracy, demonstrating the practical feasibility of implementing complex  transformations within a single compact optical element.




\section{Results and discussion}
\subsection{Spherical multipole representation}


\begin{figure}[t]
    \centering
    \includegraphics[width=\columnwidth,height=0.58\textheight,keepaspectratio]{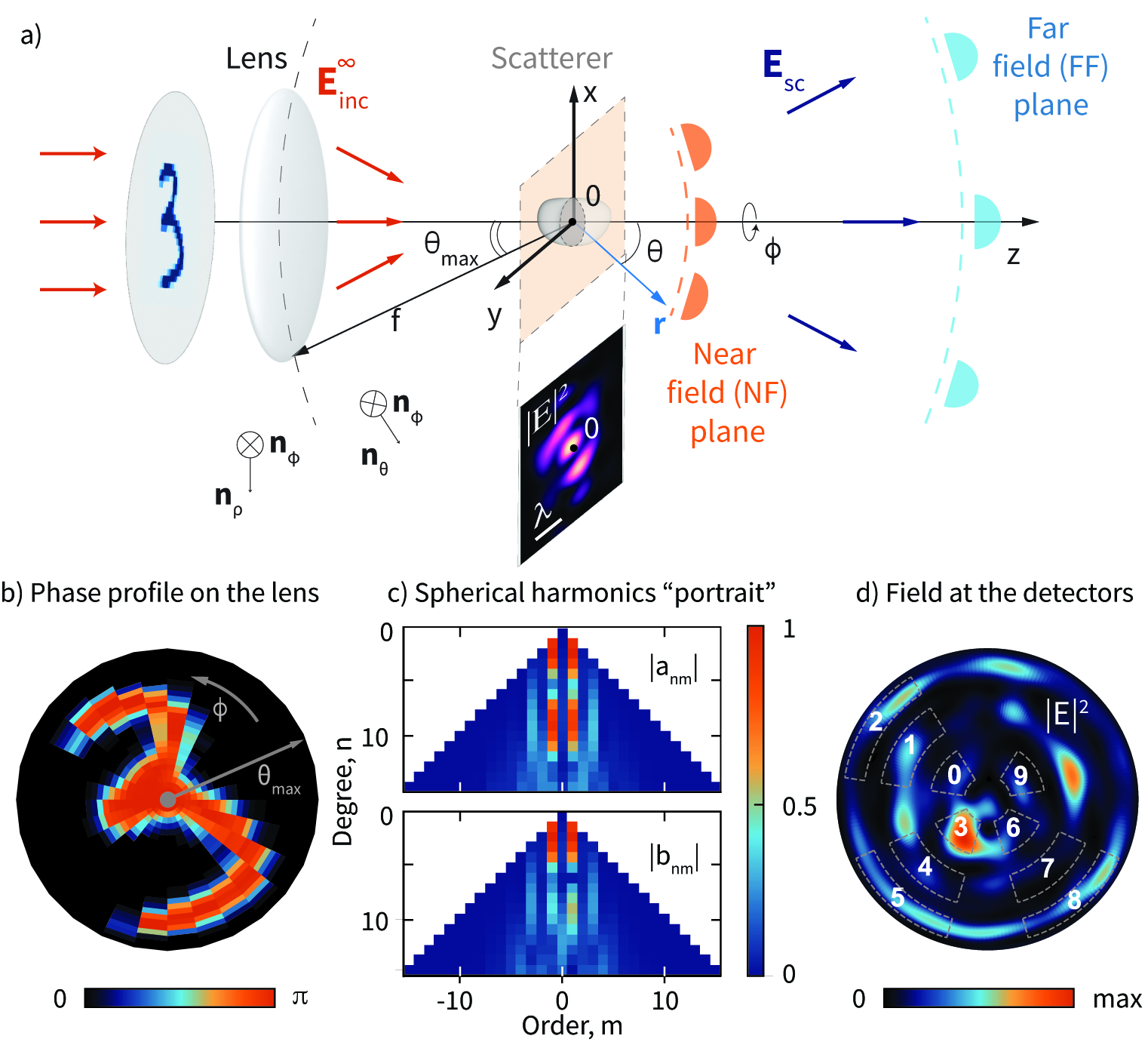}
    \caption[Mie particle  optical computing schematic ]{
        (a) Near-field and far-field formulation of the T-matrix-based optical computation scheme.
        (b) Example of an MNIST digit encoded in the spherical-harmonic modal basis. 
        The input image is encoded in phase of $x$-component of electric field projected on a lens.
        (c) Amplitudes of the multipole decomposition of the input field in (b).
        (d) Intensity pattern of scattered fields in the far-field plane. Detectors regions are denoted with dashed lines.
        }
    \label{fig:figure1}
\end{figure}

We consider a single scatterer as an optical computation unit which  performs a trainable linear transformation of the incident structured electromagnetic fields in the basis of vector spherical harmonics. The general architecture of the system is illustrated in Fig.~\ref{fig:figure1} (a).  In this system, input information is encoded in the electromagnetic wave that passes through the infinitely thin lens and focused onto the particle. The scatterer performs processing of the input and the output of the system is collected by the set of detectors. In our work we demonstrate the working principle of the computational unit on the MNIST handwritten digits classification task, where the image of the digit serves as input. Detectors collect the scattered-field intensity in the near-field zone ($kr\sim n_{\max}$) or far-field ($kr\gg 1$) of the scatterer. The  number of detectors is equal to the number of classes $K$ to make one-to-one correspondence between detector number and digit class. 

Here, we consider monochromatic fields with the wavelength $\lambda$ ($k=2\pi/\lambda$). The lens is characterized by its numerical aperture (NA) which defines the maximal angle of the incident light $\theta_{\max}=\arcsin(\mathrm{NA})$. The NA of the detector grid is equal to the NA of the lens for definiteness. The maximal illumination angle determines how many transverse spatial frequencies of the input field can reach the scatterer. Larger $\theta_{\max}$, or equivalently larger NA, excites more rapidly varying angular structure and therefore requires a larger set of multipole channels for an accurate representation. At the same time, the number of physically relevant multipoles is limited by the particle size parameter $ka$: a scatterer with larger $ka$ can support higher-order electric and magnetic multipoles. Thus, the truncation order $n_{\max}$ must be chosen large enough to resolve both the angular content imposed by the lens and the highest multipolar orders supported by the particle, with the natural scaling $n_{\max}\sim ka$ up to a numerical safety margin.



\begin{figure}[htbp]
    \centering
    \includegraphics[width=\columnwidth,height=0.58\textheight,keepaspectratio]{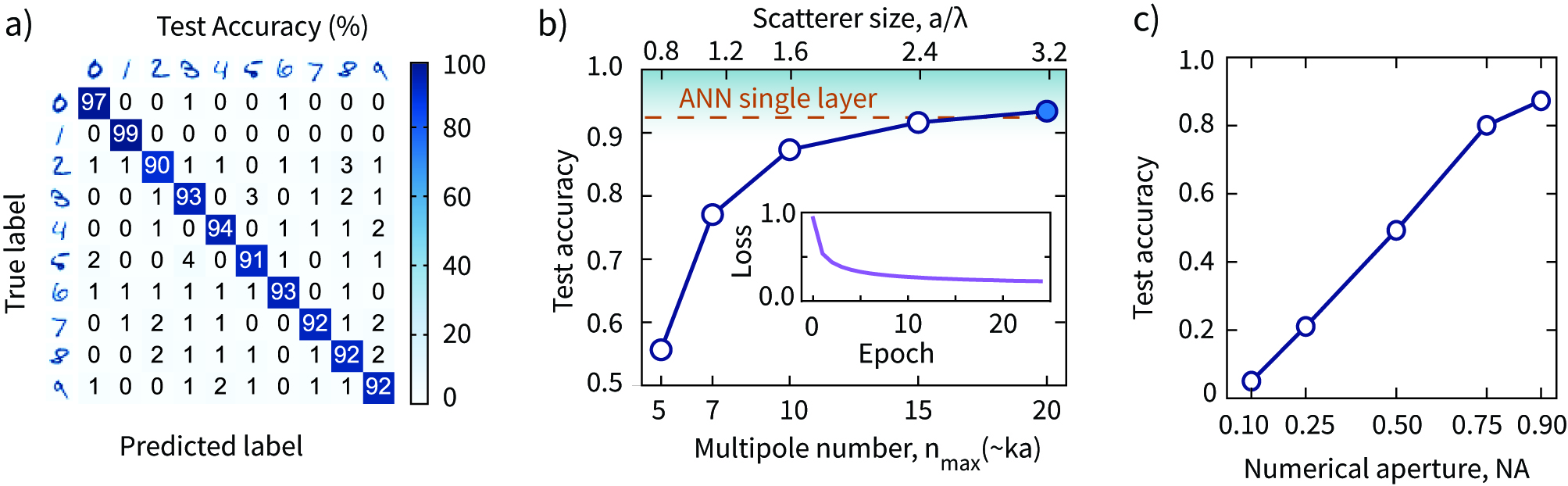}
    \caption[Classification accuracy of the T-matrix optical classifier]{
        (a)          Confusion matrix for the MNIST test set, with rows and columns denoting true and predicted labels, respectively ($n_{\max}=20$, NA$=0.9$, corresponds to blue point in (b)).         
       (b) MNIST test accuracy as a function of the maximum retained multipole order $n_{\max}$ for the T-matrix-based optical classifier (NA$=0.9$). 
        The upper axis shows the corresponding scatterer size parameter $a/\lambda$, the dashed line indicates the reference accuracy of a single-layer artificial neural network, and the inset shows the training-loss evolution for $n_{\max}=20$. 
        (c) Test accuracy as a function of the numerical aperture ($n_{\max}=10$).
    }
    \label{fig:figure2}
\end{figure}

The incident and scattered fields are expanded in the basis of vector spherical harmonics as
\begin{equation}
    \mathbf{E}_{\text{inc, sc}}(\mathbf{r}) =
    \sum_{n=1}^{n_{\max}} \sum_{m=-n}^{n}
    \left[
        a_{nm} (p_{nm})\,\mathbf{M}^{(1),(3)}_{nm}(k\mathbf{r})
        +
        b_{nm}(q_{nm})\,\mathbf{N}^{(1),(3)}_{nm}(k\mathbf{r})
    \right],
\end{equation}

where "inc" and "sc" subscripts denote the incident and scattered fields, respectively, and the superscripts $(1)$ and $(3)$ correspond to regular and outgoing electric ($\mathbf{N}_{nm}$) and magnetic ($\mathbf{M}_{nm}$) vector spherical wave functions \cite{mishchenko_scattering_2002}. The coefficients $a_{nm} (p_{nm})$ and $b_{nm}(q_{nm})$ represent the multipolar content of the incident (scattered) field. The total field is given by the sum of the incident and scattered fields:
\begin{equation}
    \mathbf{E}(\mathbf{r}) = \mathbf{E}_{\mathrm{inc}}(\mathbf{r}) + \mathbf{E}_{\mathrm{sc}}(\mathbf{r}).
\end{equation}

In our formulation, the input object is a far-field image encoded as an angular field distribution incident on the particle after lens transmission. The input image can be represented in the angular domain by a complex vector field defined over the numerical-aperture cone $\Omega_{\mathrm{NA}}$ right before the lens with $\mathbf{E}_{\mathrm{inp}}^{\infty}(\theta,\phi)$ and right after the lens with $\mathbf{E}_{\mathrm{inc}}^{\infty}(\theta,\phi)$ at the radius $r=f$, where $f$ is the focal length. The angular field is then projected onto the far-field angular parts $\boldsymbol{\mathcal{A}}_{nm}(\theta,\phi)$ and $\boldsymbol{\mathcal{B}}_{nm}(\theta,\phi)$ associated with vector spherical harmonics (see details in Methods Sec.~\ref{sec:incident_coefficients}) 
\begin{align}
    \label{eq:projection1}
    a_{nm} &= \int_{\Omega_{\mathrm{NA}}}
    \mathbf{E}_{\mathrm{inc}}^{\infty}(\theta,\phi)\cdot
    \boldsymbol{\mathcal{A}}_{nm}^{*}(\theta,\phi)\, d\Omega, \\
    \label{eq:projection2}
    b_{nm} &= \int_{\Omega_{\mathrm{NA}}}
    \mathbf{E}_{\mathrm{inc}}^{\infty}(\theta,\phi)\cdot
    \boldsymbol{\mathcal{B}}_{nm}^{*}(\theta,\phi)\, d\Omega,
\end{align}
where $d\Omega = \sin\theta\, d\theta\, d\phi$. These coefficients define the modal representation of the input image. The input image is scalar and linearly polarized, the angular field may be written as
\begin{equation}
    \mathbf{E}_{\mathrm{inp}}^{\infty}(\theta,\phi)
    =
    u(\theta,\phi)\,\hat{\mathbf{e}}_{\mathrm{pol}}(\theta,\phi),
\end{equation}
where $u(\theta,\phi)$ is the complex image amplitude and $\hat{\mathbf{e}}_{\mathrm{pol}}$ is the local polarization unit vector. The modal coefficients are then computed by substituting the corresponding field after lens transmission $\mathbf{E}_{\mathrm{inc}}^{\infty}$ into the projection formulas \eqref{eq:projection1}-\eqref{eq:projection2} above.

An example of the MNIST digit that is encoded in the far field as a linearly polarized angular field distribution and projected onto the lens reference sphere via a gnomonic projection is shown in the Figure \ref{fig:figure1}(b). The image is encoded in the phase of the $x$-component of the electric field on a grid of $28\times28$ points in $\theta$ and $\phi$ coordinates, i.e 
$ \hat{\mathbf{e}}_{\mathrm{pol}} = \mathbf{e}_x $ and  
$ u = \exp{i \pi \cdot \text{image}_{\text{proj}}}$, where $\text{image}_{\text{proj}}$ encodes  the gnomonic projection of the input image (see Methods Sec.~\ref{sec:incident_coefficients}).
The representation of the input image in the multipolar basis (spherical harmonics ``portrait") is shown in Fig.~\ref{fig:figure1}(c) for magnetic and electric amplitude coefficients. Evaluated coefficients for all the samples form the MNIST dataset in multipole represenation (see Supplementary Materials Sec.~C). One can notice that the input image is not fully captured by the first few multipole coefficients and requires modes with higher $n$. At the same time, linearly polarized excitation efficiently populates mainly modes with small $|m|$, which is inefficient in terms of input-feature dimensionality, but determines the potential capacity of the spherical multipole representation for the input data.


\subsection{Optical classification procedure}

In the proposed architecture, the trainable transformation is implemented by a single compact scatterer within the T-matrix formalism. The T-matrix describes the linear mapping between the incident and scattered multipole coefficients \cite{mishchenko_scattering_2002}, effectively encoding the computational operation of the optical processor into the modal scattering response of a compact  object: 
\begin{equation}
    \label{eq:T-matrix}
    \begin{pmatrix}
        \mathbf{p} \\
        \mathbf{q}
    \end{pmatrix}
    =
    \begin{pmatrix}
        \mathbf{T}^{11} & \mathbf{T}^{12} \\
        \mathbf{T}^{21} & \mathbf{T}^{22}
    \end{pmatrix}
    \begin{pmatrix}
        \mathbf{a} \\
        \mathbf{b}
    \end{pmatrix}.
\end{equation}
Here, $\mathbf{a}$, $\mathbf{b}$, $\mathbf{p}$, and $\mathbf{q}$ are column vectors obtained by stacking the coefficients 
$a_{nm}$, $b_{nm}$, $p_{nm}$, and $q_{nm}$, respectively. 
Since the multipole expansion is truncated at $n_{\max}$, the T-matrix generally is a finite-dimensional matrix of size $2N \times 2N$, 
where $N = n_{\max}(n_{\max}+2)$ is the number of multipole modes retained in the expansion.
Therefore, $n_{\max}$ determines not only the quality of the multipole representation of the input data but also the number of trainable parameters in the T-matrix, which scales approximately as $\sim n_{\max}^4$. This quartic scaling reflects dense pairwise coupling between two modal spaces whose individual dimensionalities scale as $n_{\max}^2$.
The relation Eq.~\eqref{eq:T-matrix} defines the central linear operation of the optical network (see  Supplementary Materials Sec.~D) and the weight matrix that can be trained to perform the classification task. 

Once the scattered coefficients are obtained, the scattered electromagnetic field is reconstructed from the multipole expansion.
In the far field, the scattered field takes the asymptotic form $\mathbf{E}^{\infty}_{\mathrm{sca}}(\theta, \phi)$ based on the asymptotic behavior of the outgoing vector spherical wave functions $\mathbf{M}_{nm}$ and $\mathbf{N}_{nm}$ ~\cite{mishchenko_scattering_2002}.

The classification signal is obtained by integrating the scattered intensity over a set of detector regions $\Omega_k$ at a radial distance $r_{\mathrm{det}}$, corresponding to either the near- or far-field detection regime:
\begin{equation}
    I_k =
    \int_{\Omega_k}
    \left|
        \mathbf{E}_{\mathrm{sc}}(r_\mathrm{det}; \theta,\phi)
    \right|^2
    d\Omega,
\end{equation}
where $k=1,\dots,K$ labels the detectors and $K$ is the number of classes, and, in the case of MNIST dataset, $K=10$.
Each detector therefore collects optical power within a prescribed angular sector.
Detectors positions and the resulting intensity in detection plane are illustrated in Figure \ref{fig:figure1}(d) for the far-field formulation where "3" digit is classified correctly with the highest intensity among all the detectors. 

Alternatively, the classification can be performed in the {\it near-field zone}, where the scattered field is evaluated at a distance  $\lambda$ from the surface of the implied scatterer and the intensity is integrated over detector regions $\Omega_k$ on a spherical surface with $kr=n_{max}+2\pi$. 

\subsection{Training procedure}
The detector outputs $\{I_k\}$ are interpreted as class scores. For a sample with ground-truth label $y$, we define the classification loss by the softmax cross-entropy
\begin{equation}\label{eq:objective}
    \mathcal{L}_{\mathrm{cls}}
    =
    -\log
    \frac{\exp(I_y)}
    {\sum_{k=1}^{K}\exp(I_k)}.
\end{equation}
The trainable parameters are the admissible elements of the T-matrix. In our work, all the constraints were implemented using parameterization and regularization of the T-matrix, so the loss function was not explicitly modified to include additional penalty terms. Details of the optimization procedure and the implementation of physical constraints are given in Methods Sec.~\ref{sec:optimization}.

\subsection{Results of training and classification performance}

Figure \ref{fig:figure2} summarizes the performance of the T-matrix-based optical classifier on the MNIST dataset. In panel Fig.~\ref{fig:figure2} (a), the confusion matrix is shown for the case $\mathrm{NA}=0.9$ and $n_{\max}=20$, corresponding to the filled blue marker in Fig.~\ref{fig:figure2}(b). It confirms that the trained T-matrix achieves strong discrimination across most digit classes, with only a few off-diagonal misclassifications remaining. Fig.~\ref{fig:figure2}(b) shows the test accuracy as a function of the retained multipole order $n_{\max}$ ($\mathrm{NA}=0.9$), showing that the classification performance improves steadily with increasing modal capacity and begins to saturate once enough multipole channels are included. The dashed reference line indicates the accuracy of a single-layer artificial neural network (ANN, see details in Methods Sec.~\ref{sec:optimization}), while the inset shows the training-loss evolution during optimization for $n_{\max}=20, \text{NA}=0.9$, indicated by the filled blue marker. Similar results were obtained for the total field formulation (see Supplementary Materials Sec.~A). The number of free parameters in ANN is approximately $7.8\cdot 10^3$ that matches well with number of free parameters in the T-matrix with $n_{\max}=15$ (approximately $5.9\cdot10^3$ with assumed reciprocity, passivity and axial symmetry - see Sec.~\ref{sec:phys_const}). Thus, a single compact scatterer described by an arbitrary T-matrix with a sufficiently rich multipolar response can perform nontrivial classification tasks on structured optical inputs.

Panel Fig.~\ref{fig:figure2} (c) reports the classification accuracy as a function of the numerical aperture for fixed  multipole order $n_{\max}=10$. The required number of multipoles is directly related to the NA of the system. Indeed,  small NA will require significantly larger number of   angular frequencies to capture the features within small $\theta_{\max}$  and processed by the scatterer. Thus, simultaneous increase of the NA and the number of multipole modes leads to improved classification performance, while insufficient modal capacity or limited NA can suppress the ability of the system to discriminate between different classes effectively.

Together, the results in Fig.~\ref{fig:figure2} indicate that both the number of retained multipole modes and the numerical aperture are key factors in achieving reliable optical classification with the T-matrix-based architecture.

\begin{figure}[htbp]
    \centering
    \includegraphics[width=0.5\columnwidth,keepaspectratio]{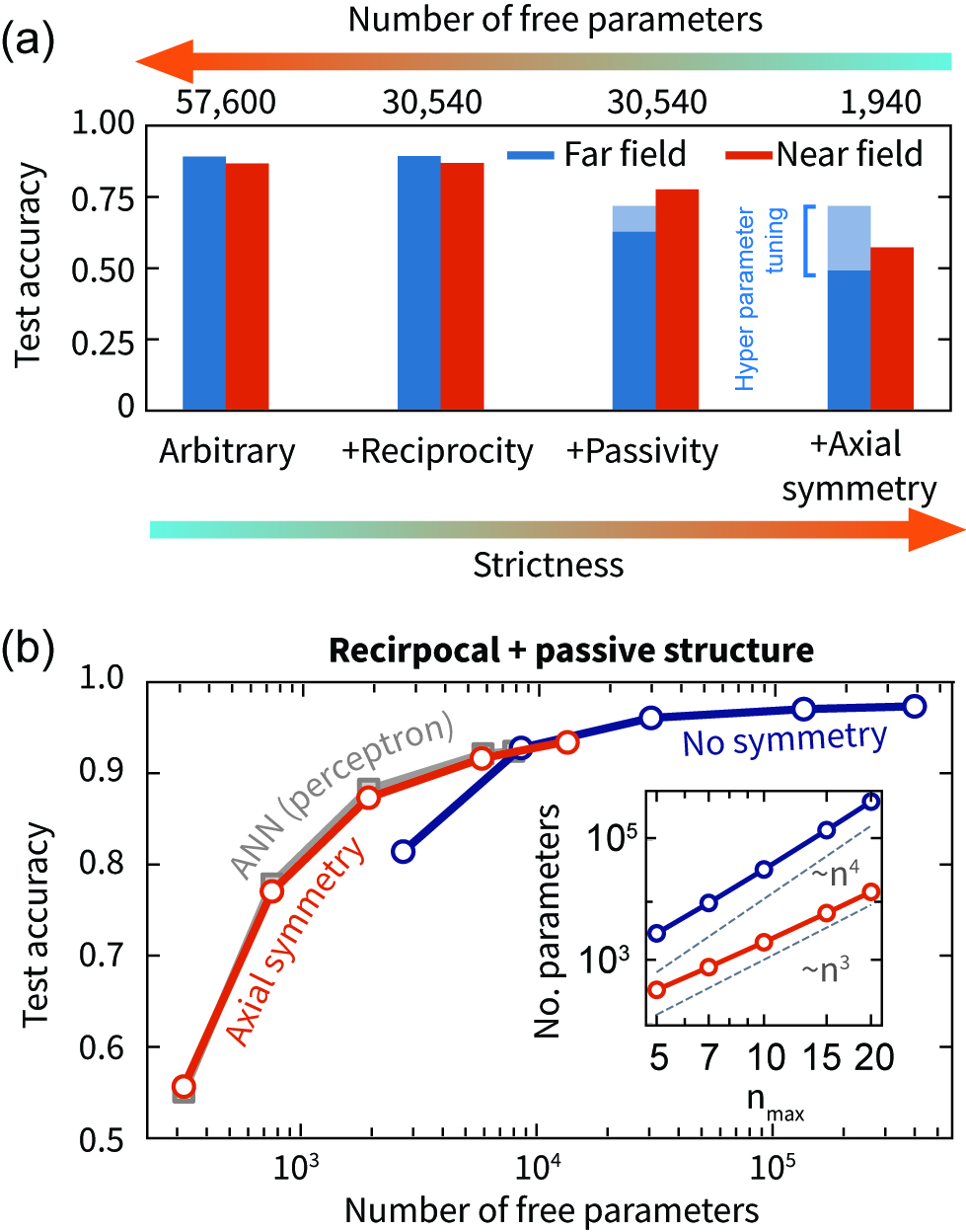}
    \caption[Mie T-matrix classifier accuracy and ANN comparison]{
        (a) Classification accuracy for reciprocal, passive, and axially symmetric T-matrices trained at $n_{\max}=10$ and $\mathrm{NA}=0.5$ in both near- and far-field formulations. The panel shows how reciprocity, passivity, and axial symmetry affect the achievable performance under electromagnetic constraints.
        (b) Comparison of the constrained T-matrix classifiers with a single-layer ANN of similar parameter count, demonstrating that the physically constrained multipolar scatterer can achieve competitive accuracy.
    }
    \label{fig:figure3}
\end{figure}

\subsection{Physical constraints}\label{sec:phys_const}
One of the central questions in the design of optical neural networks is how physical constraints on wave propagation and scattering affect the achievable performance and scalability of the system \cite{li_spatial_2025,miller_communicating_2000}. In the case of a single scatterer, unlike an unconstrained neural-network weight matrix, the T-matrix is restricted by electromagnetic principles \cite{molesky_t-operator_2022}. There are several key constraints that must be considered when designing and training the T-matrix-based optical classifier, including reciprocity, passivity, energy conservation, and symmetry relations. While the results shown in Fig.~\ref{fig:figure2} were obtained by enforcing these constraints without explicitly discussing them, it is important to understand how they can be incorporated into the T-matrix and how they affect the classification performance.

\textbf{Reciprocity.} While there are many theoretical and experimental examples of non-reciprocal single scatterers \cite{goh_nonreciprocal_2024,sounas_giant_2013,caloz_electromagnetic_2018}, in the context of passive linear optical systems, reciprocity remains a fundamental constraint in the T-matrix formalism. Reciprocity imposes symmetry relations between coupling channels \cite{mishchenko_scattering_2002}:
\begin{equation}
    T^{kl}_{n,m,n',m'}
    =
    (-1)^{m+m'}
    T^{lk}_{n',-m',n,-m},
\end{equation}
where $k,l=1,2$ denote the electric and magnetic multipole channels, respectively, and $n,m$ are the multipole indices.   When reciprocity is imposed, the number of free parameters is reduced by approximately a factor of two. For sufficiently large $n_{\max}$, the resulting accuracy can remain comparable to that of an unconstrained T-matrix because the model still retains enough trainable degrees of freedom.

\begin{figure}[htbp]
    \centering
    \includegraphics[width=\columnwidth,height=0.68\textheight,keepaspectratio]{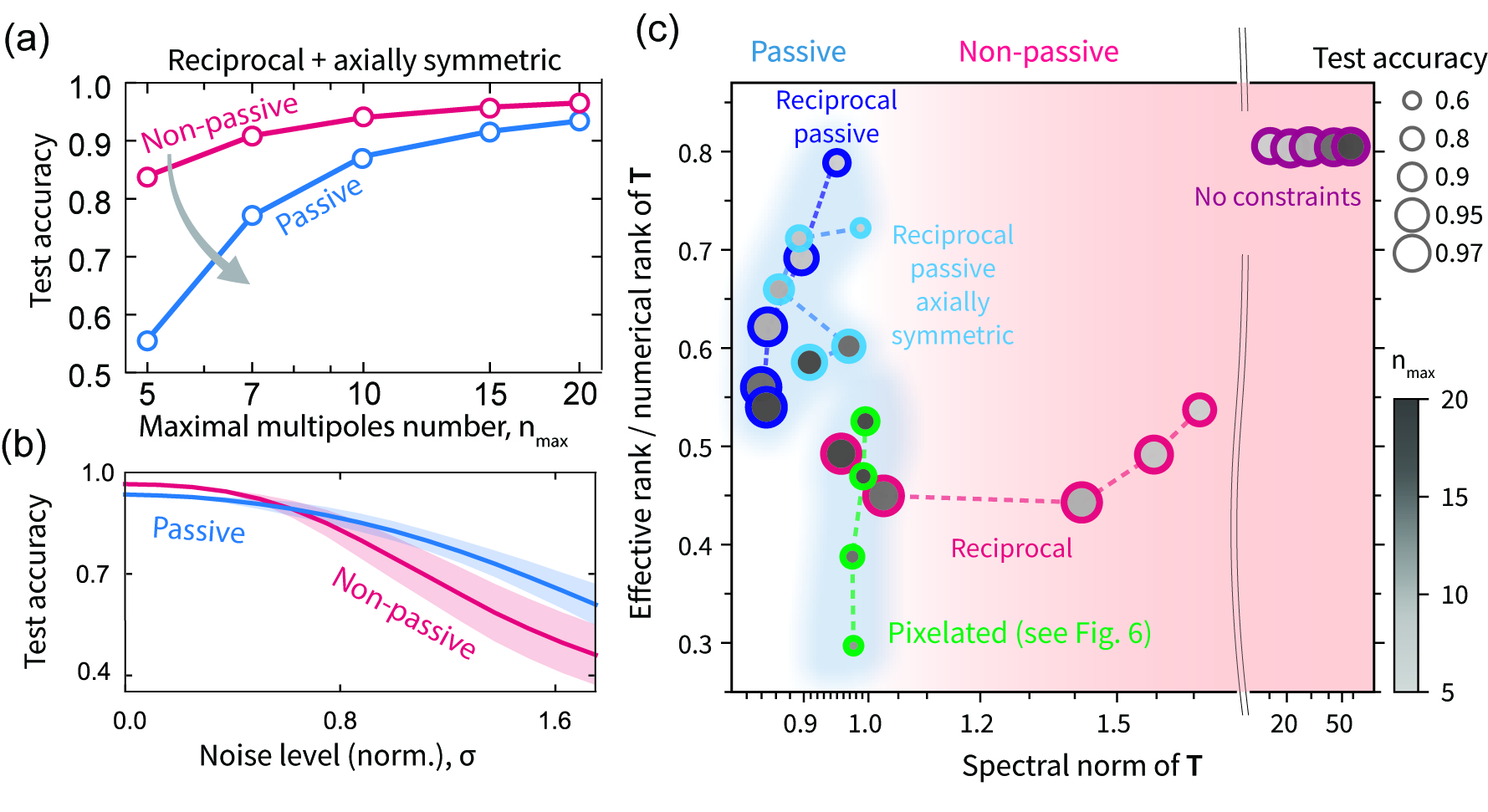}
    \caption[T-matrices passivity effects]{
        (a) Accuracy as a function of retained multipole order for the far-field formulation at $\mathrm{NA}=0.9$ with symmetrical reciprocal T-matrices, showing that higher modal capacity reduces the performance gap introduced by passivity.
        (b) Classification accuracy of the T-matrix-based classifier as a function of the noise magnitude $\sigma$ for the symmetrical reciprocal passive T-matrix and symmetrical reciprocal T-matrix for $n_{\max} = 20$ and NA $=0.9$.
        (c) Distribution of the optimized T-matrices and accuracy in the coordinate space defined by the spectral norm of the scattering matrix $\|\mathbf{T}\|_2$ and the spectral entropy rank of the T-matrix.
    }
    \label{fig:figure4}
\end{figure}

\textbf{Passivity.} Further constrains the admissible parameter space by requiring that the scattered power does not exceed the incident power. In the T-matrix formalism, equality of scattered and incident power imposes a generalized unitarity (optical theorem) condition on the transition operator. In operator form, this can be written as
\begin{equation}
    \mathbf{T} + \mathbf{T}^\dagger + 2\,\mathbf{T}^\dagger \mathbf{T} = 0.
\end{equation}
In terms of S-matrix, where $\mathbf{S}=\mathbf{E}+2\mathbf{T}$, this condition can be written as $\mathbf{S}^\dagger\mathbf{S}=\mathbf{E}$ with $\|\mathbf{S}\|_2=1$. 
The required passivity condition can be formulated as 
\begin{equation}
    \|\mathbf{S}\|_2\leq1,
\end{equation}
which ensures that the extracted power (extinction) equals the sum of scattered and absorbed power and that no net energy is generated by the system. Equivalently, in terms of modal coefficients, the passivity condition implies
\begin{equation}
    \|\mathbf{p}\|^2 + \|\mathbf{q}\|^2 \leq \|\mathbf{a}\|^2 + \|\mathbf{b}\|^2,
\end{equation}
for any  incident field amplitudes $(\mathbf{a},\mathbf{b})$, where $\|\cdot\|$ denotes the appropriate power-normalized norm of multipole coefficients. The passivity constraint does not reduce the nominal number of parameters, but it restricts optimization to the manifold of passive scatterers. 
Therefore, passivity can reduce the achievable accuracy even when the number of trainable coefficients is unchanged.

\textbf{Axial symmetry} of the scatterer can be additionally imposed by requiring that the T-matrix elements vanish for $m \neq m'$. This constraint, on the contrary to previously discussed limitations, substantially reduces the number of free parameters because the T-matrix becomes block-diagonal in $m$ and couples only modes with identical azimuthal index:
\begin{equation}
    T^{kl}_{n,m,n',m'}
    =
    \delta_{mm'}\,T^{kl}_{n,n',m,m}.
\end{equation}
This symmetry of the T-matrix leads to a reduction in the number of independent parameters and also results in $\sim n_{\max}^3$ scaling. In the following, we will rely on axially symmetric scatterer designs in order to reduce the computational complexity during the scatterer design process and to reduce the number of trainable parameters.

Thus, the nominally large modal-coupling space of the T-matrix is progressively restricted by physical constraints: reciprocity and symmetry reduce number of independent parameters, while passivity restricts the admissible operator spectrum. More generally, not every desired set of input-output transformations can be realized by a single physical structure~\cite{molesky_t-operator_2022}.

The results of the simulations with the physically constrained T-matrix are summarized in Figure \ref{fig:figure3}. Panel (a) compares classification accuracy under different physical constraints, starting from an arbitrary T-matrix and then sequentially imposing reciprocity, passivity, and axial symmetry at $n_{\max}=10$ and $\mathrm{NA}=0.5$ in both near- and far-field formulations. The corresponding number of free parameters is shown above each case. Reciprocity alone has only a mild impact on performance, whereas passivity suddenly causes a clear accuracy decrease despite leaving the nominal parameter count unchanged. This indicates that the main effect of passivity is not parameter reduction, but the spectral restriction imposed on the scattering operator: the classifier can no longer use arbitrary amplification or non-passive mode mixing to separate classes.

The scaling of the accuracy with the number of trainable parameters is shown in Figure~\ref{fig:figure3}~(b) for the case of reciprocal and passive structure (see results for the other constraints combinations in Supplementary Materials Sec.~B). The blue line corresponds to the reciprocal passive system, while the orange line corresponds to the axially symmetric case. In both cases, the multipole order $n_{\max}$ is varied, so that the number of free parameters changes together with the modal capacity of the classifier. Because axial symmetry conserves the azimuthal index $m$, the axially symmetric T-matrix contains approximately one order of magnitude fewer trainable parameters for the same $n_{\max}$. Nevertheless, its accuracy scales comparably to that of the ANN baseline and can remain competitive even with a substantially smaller parameter count. Here, the single-layer ANN was implemented as a linear classifier and trained similarly using softmax cross-entropy loss and masking out a random subset of weights to investigate dependence on the parameters number. The inset of Figure \ref{fig:figure3} (b) shows the corresponding scaling of the T-matrix number of parameters with $n_{\max}$.

The  effect of accuracy drop due to passivity, in our opinion, deserves a more detailed discussion.  Figure \ref{fig:figure4} (a) shows the accuracy as a function of the retained multipole order in the far-field formulation for reciprocal structures with axial symmetry. The comparison highlights that passivity leads to a significant drop in accuracy, although the number of trainable parameters is the same as in the reciprocal non-passive case. 
This behavior highlights the role of passivity as a spectral constraint and indicate that the performance loss cannot be explained only by the parameter count. 

In the non-passive model, training can exploit large singular values of the T-matrix to amplify class-separating directions in the multipolar feature space. In the passive model, the associated scattering matrix must satisfy $\|\mathbf{S}\|_2 \leq 1$, so the classifier cannot increase the optical power in arbitrary outgoing channels. This distinction also suggests that the passive model may be more sensitive to the details of the training procedure. 
Moreover, this bound limits the scale of the detector intensities and, consequently, the classification margins for a fixed incident field. This is particularly relevant for physical neural networks, where the squared field amplitudes at the detectors can be small. In this case, the resulting bound on the cross-entropy loss and the reduced gradients can make it difficult to separate samples across the decision boundary during finite-time optimization (see  Supplementary Materials Sec.~E).
Related studies on Lipschitz-constrained networks with spectral normalization report competitive accuracy and, in some cases, improved robustness \cite{bethune_pay_2022}. In practice, however, achieving such performance requires careful training design, including appropriate dataset normalization to produce logits with a suitable scale for cross-entropy optimization. 
Alternatively, this scale can be controlled by introducing a softmax temperature $\tau$ \cite{agarwala_temperature_2020}. The loss depends on the classification margins $m$ through $m/\tau$, so choosing $\tau<1$ increases the effective margin and strengthens the training signal from a given physical intensity difference. Temperature can therefore partially compensate for the optimization penalty imposed by bounded optical gain.

At the same time, bounded gain suppresses the amplification of perturbations. To test this regularization effect directly, we add complex Gaussian perturbations to the trained T-matrix during inference,
$$
    \mathbf{T}_{\text{noisy}} = \mathbf{T} + \sigma \cdot \gamma \cdot \mathbf{N}, \quad \mathbf{N} \sim \mathcal{CN}(0, 1)
$$
where $\sigma$ is the dimensionless noise magnitude and $\mathbf{N}$ is a matrix of independent and identically distributed complex Gaussian samples. The scale factor $\gamma=\frac{\|T\|_F}{\sqrt{N}}$ is chosen from the normalized Frobenius norm of the trained matrix, so that the perturbation amplitude is proportional to the typical magnitude of the T-matrix elements. 
Figure~\ref{fig:figure4} (b) compares axially symmetric reciprocal T-matrices trained with and without passivity at $n_{\max}=20$ and $\mathrm{NA}=0.9$.
It is observed that the passive T-matrix model exhibits stronger robustness to noise compared to the unconstrained model, which can be attributed to the regularizing effect of the passivity constraint that limits the spectral norm of the T-matrix and prevents excessive amplification of perturbations.

Finally, in Fig.~\ref{fig:figure4} (c) we summarized all optimized T-matrix solutions in the coordinates of the spectral norm $\|\mathbf{T}\|_2$ and the normalized effective rank
\begin{equation}
    r_{\mathrm{eff}} = \frac{
        \exp\left(-\sum_i \tilde{s}_i \log \tilde{s}_i\right)
    }{
        \mathrm{rank}(\mathbf{T})
    }, \qquad
    \tilde{s}_i=\frac{s_i}{\sum_j s_j},
\end{equation}
with $s_i$ being the singular values of $\mathbf{T}$. The quantity $r_{\mathrm{eff}}$ estimates the fraction of singular scattering channels that are effectively used by the trained operator. Values close to unity correspond to a broad singular-value spectrum, where many channels contribute with comparable weights, while lower values indicate that the transformation is dominated by a smaller number of relatively high-gain channels.
This quantity shows the distinction between nominal parameter count and effectively utilized scattering channels demonstrating that large T-matrix does not by itself imply a proportionally large information capacity~\cite{amaolo_maximum_2026}.
The spectral norm of $\mathbf{T}$ is not itself a complete passivity criterion, because passivity is imposed on $\mathbf{S}=\mathbf{I}+2\mathbf{T}$ through $\|\mathbf{S}\|_2\leq 1$. Nevertheless, $\|\mathbf{T}\|_2>1$ is incompatible with passivity, while $\|\mathbf{T}\|_2\leq 1$ is only a necessary, not sufficient, indicator of passive behavior.

In all the types of solutions accuracy (shown with the circle size in Fig.~\ref{fig:figure4}(c)) grows with number of multipoles from $n_{\max}=5$ to $n_{\max}=20$ (shown with the grey filling color), as was already discussed previously, and reaches high accuracy even with small number of multipole modes used for the unconstrained and reciprocal non-passive solutions. These solutions occupy the larger $\|\mathbf{T}\|_2$-values, while passive solutions are confined to the low-norm region. The effective-rank trend shows that non-constrained models use all the available channels, while passive and axially symmetric passive classifiers tend to use relatively fewer effective singular channels as the modal basis grows. 

To recapitulate, Fig.~\ref{fig:figure4} identifies two distinct mechanisms: non-constrained models improve classification by channels amplification, whereas passive models trade part of this accuracy for bounded gain, physical admissibility, and enhanced robustness to parameter noise.

\subsection{Designing a physical scatterer}

\begin{figure}[htbp]
    \centering
    \includegraphics[width=\columnwidth,height=0.68\textheight,keepaspectratio]{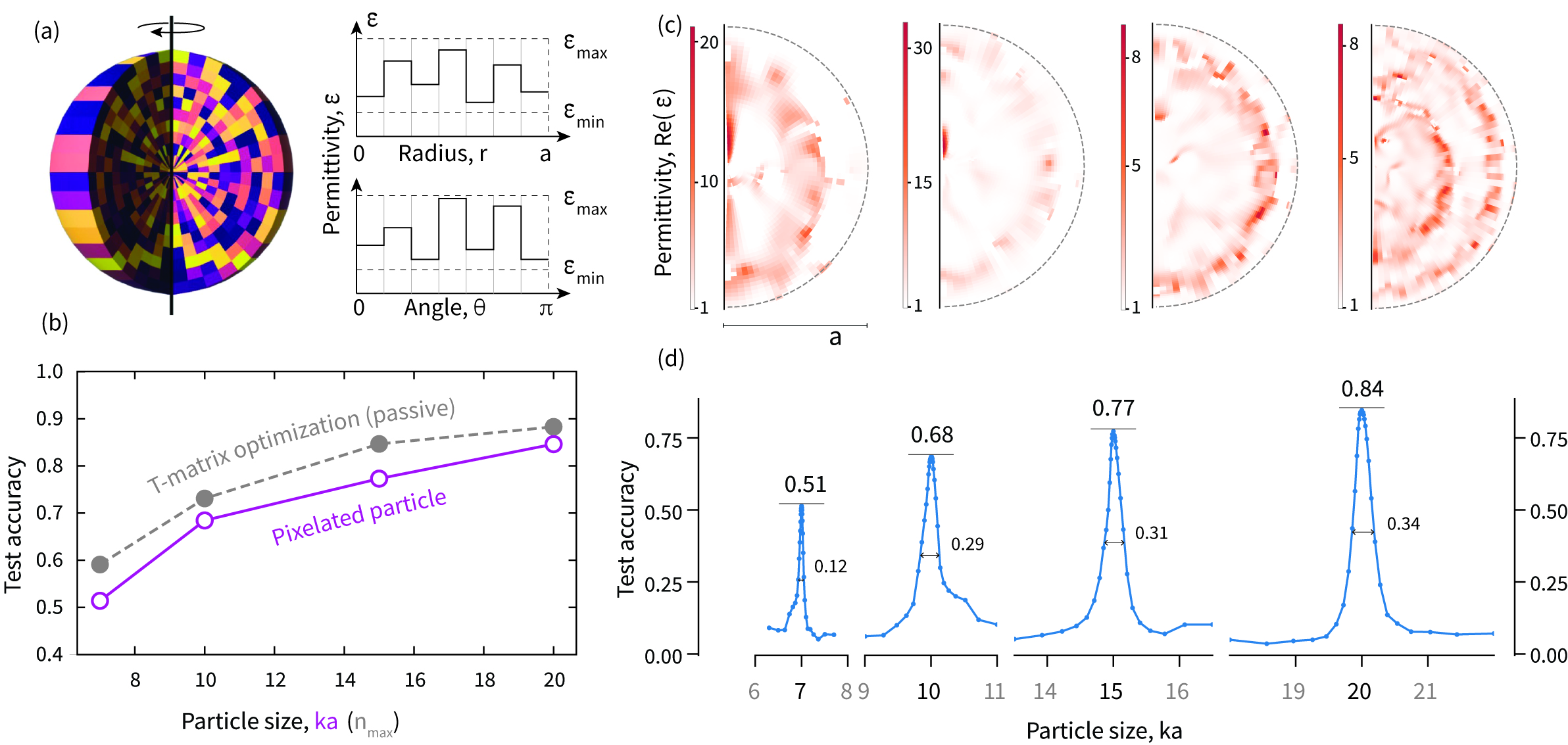}
    \caption[Physically realizable pixelated scatterer]{
        (a) Parameterization of an axially symmetric pixelated particle with a trainable relative-permittivity distribution discretized along the radial coordinate $r$ and polar angle $\theta$, with each pixel constrained to $\varepsilon_{\min}\leq\varepsilon\leq\varepsilon_{\max}$.
        (b) Classification accuracy of the optimized pixelated particle compared with the reciprocal, passive, and axially symmetric T-matrix as a function of the particle size parameter $ka$ and the associated multipole truncation order $n_{\max}$.
        (c) Optimized permittivity distributions obtained for representative particle sizes.
        (d) Test accuracy as a function of $ka$ in the vicinity of the selected resonant solutions. The horizontal markers and numerical values indicate the maximum accuracy achieved near $ka=7$, $10$, $15$, and $20$, while the horizontal arrows and the corresponding numerical values denote the full width at half maximum (FWHM) of each resonance.
    }
    \label{fig:figure5}
\end{figure}

The results above establish that reciprocity, passivity, and axial symmetry still leave sufficient modal capacity for nontrivial optical classification. However, these constraints are only necessary conditions for physical realizability: an arbitrary T-matrix satisfying them does not necessarily correspond to a scatterer with a realizable spatial distribution of material parameters. 
In particular, a large number of T-matrix coefficients does not imply an equal number of independently controllable material parameters, which simultaneously affect all modal couplings.
We therefore proceed from the operator-level description to a direct design of an inhomogeneous dielectric particle and examine how closely such a structure can approach the performance of the optimized physically constrained T-matrix.

The particle design and the corresponding classification results are summarized in Fig.~\ref{fig:figure5}. Here we restricted the particle geometry to be axially symmetric that reduces the number of trainable parameters and simplifies the optimization. As shown in Fig.~\ref{fig:figure5} (a), the axially symmetric particle is discretized into pixels in the radial and polar directions. Each pixel is assigned a trainable relative permittivity bounded by $\varepsilon_{\min}$ and $\varepsilon_{\max}$. The absorption losses are set to zero for this particular case so $\operatorname{Im}\varepsilon=0$ and $\operatorname{Re}\varepsilon\geq1$. Details of the optimization are given in Methods Sec.~\ref{sec:optimization}.  

The T-matrix of the pixelated particle is computed using the invariant imbedding T-matrix method (IITMM), with the computational procedure and relevant equations described in Methods Sec.~\ref{sec:iitmm}.
IITMM is an approach for calculating the T-matrices of inhomogeneous particles.
The method is particularly efficient for axially symmetric structures, providing a significant computational speedup in such cases\cite{johnson_invariant_1988,sun_invariant_2020,zhang_vector_2022}.

Figure~\ref{fig:figure5} (b) compares the classification accuracy of the optimized pixelated particle with that of the arbitrary T-matrix with reciprocal, passive, and axially symmetry restrictions (axis of symmetry is aligned with the $z$-axis). The horizontal axis relates the particle size parameter $ka$ to the corresponding multipole truncation order $n_{\max}$ ($n_{\max}\approx ka$). The pixelated realization follows the same general trend as the ideal constrained operator: the accuracy increases as the particle becomes larger and more multipolar channels become available. Although the directly optimized material distribution performs slightly below the T-matrix model over the considered range, the relatively small gap demonstrates that a substantial fraction of the operator-level performance can be retained in a spatially realizable scatterer.  In these simulations the number of pixels was set to $N_r=ka\cdot5$ and $N_\theta=60$ in the radial and polar directions, respectively, which was found to be sufficient to achieve convergence of the classification accuracy. For each value of $ka$, the T-matrix was computed with a truncation order of $n_{\max}=ka+5$ multipoles, which provides sufficient evaluation accuracy and accounts for the exponentially decaying amplitudes of the matrix elements.
  
Examples of the optimized permittivity distributions for several particle sizes are presented in Fig.~\ref{fig:figure5} (c). The resulting structures are strongly inhomogeneous and exploit both radial and angular variations of the permittivity, indicating that classification relies on coordinated coupling among multiple resonant channels rather than on a single conventional Mie resonance. Figure~\ref{fig:figure5} (d) further shows that the test accuracy depends nonmonotonically on the particle size and exhibits narrow resonant maxima near $ka=7$, $10$, $15$, and $20$ and the accuracy increases from approximately $0.51$ to $0.84$ across these parameters. At the same time,  high classification accuracy is obtained only in the very narrow spectral range  of around $\delta \lambda/\lambda\approx 2.1\cdot 10^{-2}$ when the material distribution and the size parameter jointly realize a favorable multipolar response. Out of this band, the accuracy drops to insufficient  values. 

Although the finely pixelated geometry considered here would be difficult to fabricate and implement directly at optical or microwave frequencies, it is not intended as a fabrication-ready design. Rather, it serves as a proof of principle that a T-matrix with the functionality required for classification can arise from a finite passive object with realistic material parameters. Advances in two-photon polymerization have enabled the fabrication of complex three-dimensional photonic structures with subwavelength scale features~\cite{dinc20233d,sharma20263d}, including inverse-designed volumetric scattering systems~\cite{luo2026volumetric}. In this context, the target modal response could, in principle, be approximated using more fabrication-oriented geometries, including composite scatterer arrangements optimized within the T-matrix framework~\cite{asadova_gradient-based_2026, asadova_t-matrix_2026}, multilayer and core-shell particles~\cite{peurifoy_nanophotonic_2018} and inverse-designed resonators with tailored multipolar responses~\cite{zhang_spherical-harmonic-based_2025, bahmani_topology_2025} or voxels~\cite{sved_inverse-designed_2026}.

\section{Conclusions and outlook}

An advantage of the proposed architecture is the concentration of a dense many-to-many modal transformation within a single finite scatterer with the density of free parameters overcoming the conventional layered diffractive elements by the factor of $\sim(ka)^2$ which in the case of a scatterer comparable with wavelength can give increase in two orders of magnitude. While limited density of parameters layered structures can be potentially overcome by non-locally coupled metasurfaces~\cite{overvig_nonlocal_2025}, the single scatterer immediately provides an efficient solution. This does not resolve the issue of the limited communication channel as the channel space scales as $\sim(ka)^2$ \cite{miller_tunnelling_2025}. Nevertheless, one of the big advantage of the proposed geometry is that all the computation can be performed within a near-field domain, where the number of channels can be significantly larger than in the far-field. That opens a route towards the optical computation at the interface with the integrated photonic systems allowing to directly couple preprocessed optical data to on-chip photonic circuitry.  

Along the same line, one should note that the concentration of the trainable transformation in a single scatterer does not remove the longitudinal-space requirement of the complete optical system. The phase-encoded image must first be prepared over a finite aperture and focused onto the particle, so propagation between the image plane or spatial light modulator and the scatterer remains necessary. Likewise, far-field readout requires collection optics or a propagation distance, although the near-field formulation can replace this output path by direct coupling to detectors or an integrated photonic circuit. Thus, the present architecture primarily reduces the volume occupied by the trainable transformation itself; it does not evade the diffraction- and communication-based thickness bounds that apply to the entire input--processor--output system~\cite{miller_why_2023}.


To conclude, in this work we have demonstrated that a single compact scatterer can be trained to perform nontrivial optical classification tasks on structured inputs. The T-matrix formalism provides a natural framework for describing the trainable transformation and for enforcing physical constraints, including reciprocity, passivity, and symmetry. The results show that even under these constraints, the T-matrix retains sufficient modal capacity to achieve competitive classification accuracy. Furthermore, we have shown that a physically realizable inhomogeneous dielectric particle can approximate the performance of the optimized T-matrix, demonstrating the feasibility of implementing such optical classifiers in practice. This work opens new avenues for designing compact and efficient optical neural networks based on single-scatterer architectures.

\newpage
\section{Methods}
\label{sec:Methods}

\input{Methods}

\section*{Acknowledgements}
The authors thank Andrey Bogdanov and Alexey Shcherbakov. 


\section*{Funding}

The work was supported  by the Federal Academic Leadership Program Priority 2030.

\section*{Author contributions}

V.K. contributed to methodology, software, formal analysis, investigation, visualization, conceptualization and writing of the
original draft. 
V.I. contributed to methodology, software, formal analysis, investigation, visualization, conceptualization, and writing of the original draft. 
C.D.A. contributed to conceptualization, writing -- review and editing.
M.P. contributed to conceptualization, methodology, software, formal analysis, visualization, writing of the
original draft, and supervision.
All authors discussed the results and reviewed the manuscript.

\section*{Competing interests}

The authors declare no competing interests.

\section*{Data availability}

The data that support the findings of this study are available from the corresponding author upon reasonable request.

\section*{Code availability}

The code that supports the findings of this study is available from the corresponding author upon  request.

\bibliographystyle{unsrt}
\bibliography{references}

\end{document}

%% file: Methods.tex
\subsection{Incident multipole coefficients derivation}\label{sec:incident_coefficients}

Transmitted through the ideal lens field can be expressed ~\cite{novotny_principles_2012} as
\begin{equation}
    \mathbf{E}^{\infty}_{\text{inc}}(\theta, \phi) = \left[ 
        (\mathbf{E}^{\infty}_{\text{inp}} \cdot \mathbf{n}_{\phi}) \mathbf{n}_{\phi} + 
        (\mathbf{E}^{\infty}_{\text{inp}} \cdot \mathbf{n}_{\rho}) \mathbf{n}_{\theta} 
    \right] (\cos \theta)^{1/2},
\end{equation}
where $\mathbf{n}_{\phi} = (-\sin{\phi}, \cos{\phi}, 0)$, $\mathbf{n}_{\rho} = (\cos{\phi}, \sin{\phi}, 0)$, and $\mathbf{E}^{\infty}_{\text{inp}}$ is the encoded field in the input plane, 
that is in our case is the gnomonic projection of the image on the lens (see Figure~\ref{fig:figure1}) encoded in the phase of only nonzero $x$-component of the field:
$$
    \text{image}_{\text{proj}}(\theta, \phi) = \text{image} \left( 
        \frac{\tan(\pi - \theta)}{\tan \theta_{\text{max}}} \begin{pmatrix} -\sin \phi \\ \cos \phi \end{pmatrix} 
    \right),
$$
where ${\text{image}}$ is the input image given in cartesian coordinates, and $\Theta_{\text{max}}$ is the maximum angle defined by the numerical aperture of the lens, and:
$$
    \mathbf{E}^{\infty}_{\text{inp}} = \mathbf{e}_x u(\theta, \phi), \quad 
    u(\theta, \phi) = \exp{i \pi \text{image}_{\text{proj}}}.
$$

The angular spectrum representation of the focal field is given by the angular representation of the far field on the lens reference sphere ~\cite{novotny_principles_2012}:
$$
    \mathbf{E}_{\text{inc}} = \gamma \int_0^{\theta_{\max}} \!\!\!\! \int_0^{2\pi} 
    \mathbf{E}^{\infty}_{\text{inc}}(\theta, \phi) \text{PW}(\theta, \phi) 
    d\Omega,
$$
where $d\Omega=\sin \theta \, d\phi \, d\theta$, 
$\theta_{max}$ is the maximum angle defined by the numerical aperture of the lens, 
$\gamma = -\frac{ikfe^{-ikf}}{2\pi}$, where $f$ is focal length and 
$\text{PW}(\theta, \phi) = e^{ikz \cos \theta+ik\rho \sin \theta \cos(\phi - \varphi)} = 
e^{ik_x x + ik_y y + ik_z z}$.

Since the multipole coefficients $[a/b]^{\text{PW}}_{mn}(\mathbf{E}^{\infty}_{\text{inc}},\theta, \phi)$ are known for plane waves~\cite{mishchenko_scattering_2002}:
\begin{gather*}
    a^{\text{PW}}_{mn}(\theta, \phi) = i \beta_{m,n} \mathbf{E}^{\infty}_{\text{inc}}(\theta, \phi) \cdot \mathbf{C}^*_{mn}(\theta) e^{-im\phi}, \\
    b^{\text{PW}}_{mn}(\theta, \phi) = \beta_{m,n} \mathbf{E}^{\infty}_{\text{inc}}(\theta, \phi) \cdot \mathbf{B}^*_{mn}(\theta) e^{-im\phi},
\end{gather*}
with $\beta_{m,n} = 4\pi (-1)^mi^{n-1} \sqrt{(2n+1)/(4\pi n(n+1))}$ and 
$\mathbf{C}_{mn}(\theta)$ and $\mathbf{B}_{mn}(\theta)$ are the angular parts of the magnetic and electric multipoles, respectively ~\cite{mishchenko_scattering_2002}.
the incident field can be expressed through the multipole expansion 
$$
    \mathbf{E}_{\text{inc}} = \gamma \iint \sum_{m, n} [
        a^{\text{PW}}_{mn} \mathbf{M}^{(1)}_{mn} + 
        b^{\text{PW}}_{mn} \mathbf{N}^{(1)}_{mn}
    ] d\Omega,
$$
where $\mathbf{M}^{(1)}_{mn}$ and $\mathbf{N}^{(1)}_{mn}$ are the regular vector spherical wave functions ~\cite{mishchenko_scattering_2002}
Finally, changing the order of summation and integration gives multipole decomposition of focused incident field
$$
    \mathbf{E}_{\text{inc}} = 
    \sum_{n=1}^{\infty} \sum_{m=-n}^{n} \left[ 
        a_{mn} \mathbf{M}^{(1)}_{mn}(k \mathbf{r}) + 
        b_{mn} \mathbf{N}^{(1)}_{mn}(k \mathbf{r}) 
    \right]
$$
with integral expressions for the coefficients $[a/b]_{mn}$ given by
$$
    [a/b]_{mn} = \gamma \int_0^{\theta_{\max}} \!\!\!\! \int_0^{2\pi} 
    [a/b]^{\text{PW}}_{mn}(\theta, \phi) d\Omega.
$$

This yields the final expressions for Eqs.~\eqref{eq:projection1} and~\eqref{eq:projection2}:
\begin{gather*}
    \boldsymbol{\mathcal{A}}^*_{mn}(\theta, \phi) = i \gamma \beta_{m,n} \mathbf{C}^*_{mn}(\theta) e^{-im\phi}, \\
    \boldsymbol{\mathcal{B}}^*_{mn}(\theta, \phi) = \gamma \beta_{m,n} \mathbf{B}^*_{mn}(\theta) e^{-im\phi}.
\end{gather*}

\subsection{Density of parameters}
\label{sec:dop}

For a fixed reference area $A$, taken as the projected area of the scatterer and the physical area of the diffractive layer, we estimate the scaling of the number and density of parameters for a T-matrix representation and a diffractive layer. 
Here, $a$ is the particle radius, $k$ is the wavenumber, and $\lambda$ is the wavelength.

\subsubsection{Parameter estimate for the T-matrix}

The number of parameters of an unconstrained T-matrix (modal coupling coefficients) is determined by the matrix dimension, which is set by the maximum retained multipole order $n_{\max}$:
$$
    N_{T} \sim [2 n_{\max}(n_{\max}+2)]^2 \sim 4 n_{\max}^4
$$
The relevant characteristic length scale is the particle radius $a$, with the minimal scaling requirement:
$$
    n_{\max} \sim ka
$$
The parameter count therefore has the following asymptotic scaling with $ka$:
$$
    N_{T} \sim 4 (ka)^4
$$
For a fixed area $A=\pi a^2$, the T-matrix parameter count can be written as:
$$
    N_{T} \sim 4 \left(\frac{2 \pi a}{\lambda}\right)^4 \sim 64 \pi^4 \left(\frac{a^2}{\lambda^2}\right)^2 \sim 64 \pi^2 \frac{A^2}{\lambda^4}
$$
The corresponding volumetric parameter density of the T-matrix representation is:
$$
    \rho_{T} \sim \frac{N_{T}}{(4/3) \pi a^3} \sim \frac{3}{\pi a^3} (ka)^4 \sim \frac{48 \pi^3 a}{\lambda^4}
$$

\subsubsection{Parameter estimate for a diffractive layer}

The characteristic length scale of a diffractive layer is its period $d$, taken as:
$$
d \sim \frac{\lambda}2 \cdot \frac{1}{\alpha}
$$
Here, $\alpha$ is a dimensionless factor introduced to describe metasurfaces for which $d$ can be smaller than $\lambda/2$. The number of parameters of the diffractive layer is then:
$$
    N_{dif} \sim \frac{A}{d^2} \sim \frac{4 A}{\lambda^2} \alpha^2
$$
In terms of $ka$ (using $A=\pi a^2$), the parameter count has the asymptotic scaling:
$$
    N_{dif} \sim \frac{4 \pi a^2}{\lambda^2} \alpha^2 \sim \frac{\alpha^2}{\pi} (ka)^2
$$
The corresponding areal parameter density of the diffractive layer is:
$$
    \rho_{dif} \sim \frac{N_{dif}}{A} \sim \frac{4}{\lambda^2} \alpha^2
$$

\subsection{IITMM}\label{sec:iitmm}

We calculate the T-matrix of the inhomogeneous particle using the invariant imbedding T-matrix method (IITMM)~\cite{sun_invariant_2020}. In the IITMM formulation, the radial evolution of the T-matrix is determined by the matrix $U_{mm'nn'}(r)$, whose elements contain angular integrals over the particle permittivity distribution at each radius $r$. For a pixelated particle with piecewise-constant angular permittivity, these angular integrals can be evaluated analytically, avoiding their numerical evaluation at every radial step. Following the approach of Ref.~\cite{Shcherbakov2019Jul}, we use a Legendre expansion of the angular permittivity profile.

Specifically, we write
\begin{align*}
u(r,\theta)
&= \varepsilon(r,\theta)-1
= \sum_{p=0}^{\infty}u_p(r)P_p(\cos\theta), \\
\frac{u(r,\theta)}{\varepsilon(r,\theta)}
&= \sum_{p=0}^{\infty}\widetilde{u}_p(r)P_p(\cos\theta),
\end{align*}
where $P_p$ denotes the Legendre polynomial of order $p$.

For an axially symmetric particle, the resulting matrix has the block form
\begin{equation*}
U_{mm'nn'}(r) = \frac{k^2r^2}{2} \delta_{mm'} \begin{pmatrix}
U^{11}_{mnn'}(r) & -U^{12}_{mnn'}(r) & 0 \\
U^{12}_{mnn'}(r) & U^{11}_{mnn'}(r) & 0 \\
0 & 0 & U^{33}_{mnn'}(r) \end{pmatrix}.
\end{equation*}
Its nonzero components can be evaluated directly from the Legendre coefficients as
\begin{align*}
U^{11}_{mnn'} &= -2(-1)^m\gamma_{nn'} \sum_{p=0}^{\infty} u_p
\left[1-\bigl(n+n'+p \bmod 2\bigr)\right]
C^{n p n'}_{m\,0\,-m} C^{n p n'}_{1\,0\,-1},
\\
U^{12}_{mnn'} &= -2i(-1)^m\gamma_{nn'} \sum_{p=0}^{\infty} u_p
\bigl(n+n'+p \bmod 2\bigr)
C^{n p n'}_{m\,0\,-m} C^{n p n'}_{1\,0\,-1},
\\
U^{33}_{mnn'} &= 2(-1)^m\gamma_{nn'} \sum_{p=0}^{\infty} \widetilde{u}_p
C^{n p n'}_{0\,0\,0} C^{n p n'}_{m\,0\,-m}.
\end{align*}
where $\gamma_{nn'}=\sqrt{(2n+1)(2n'+1)}$ and $C^{j_1j_2j_3}_{m_1m_2m_3}$ denotes the corresponding Wigner $3j$-symbol.

For a single angular pixel occupying
$\theta_1<\theta<\theta_2$ and having relative permittivity $\varepsilon$, its contribution to the Legendre coefficients is
\begin{equation}
u_p=(\varepsilon-1)w_p, \qquad \widetilde{u}_p=\left(1-\frac{1}{\varepsilon}\right)w_p,
\end{equation}
where
\begin{equation}
w_p= \begin{cases}
\dfrac{1}{2} \left[ P_0(\cos\theta_1)-P_0(\cos\theta_2)\right], & p=0,\\
\dfrac{1}{2} \left[
P_{p-1}(\cos\theta_2)
-P_{p+1}(\cos\theta_2)
-P_{p-1}(\cos\theta_1)
+P_{p+1}(\cos\theta_1)
\right],
& p\geq1.
\end{cases}
\end{equation}
Consequently, the angular integrals required by the IITMM can be obtained analytically for each pixel and assembled linearly for an arbitrary piecewise-constant angular permittivity profile.

\subsection{Optimization procedure}
\label{sec:optimization}

\paragraph*{T-matrix optimization.}
The T-matrix is optimized directly with respect to the classification objective of Eq.~\eqref{eq:objective}, starting from a random complex-valued initialization.
For each batch, the incident multipole coefficients are transformed $(\mathbf{a},\mathbf{b})$ by the current T-matrix according to Eq.~\eqref{eq:T-matrix}, and the resulting detector powers are used as class scores. The cross-entropy loss is differentiated with respect to the trainable complex-valued T-matrix parameters, which are updated using AdamW gradient-based optimizer.
The standard MNIST training set is divided into $55\,000$ training and $5\,000$ validation samples, while the official $10\,000$-sample test set is used only for the final evaluation.

Different physical constraints are incorporated into the same optimization procedure by restricting the admissible T-matrix rather than modifying the classification loss.
For the unconstrained model, the independent complex valued matrix elements are optimized directly.
Reciprocity and geometrical symmetries are imposed through the parameterization of the T-matrix, such that only the independent matrix elements are treated as trainable variables and the remaining elements are reconstructed from the corresponding symmetry relations.
Passivity is enforced during optimization by projecting the reconstructed scattering matrix $\mathbf{S}$ onto the unit spectral norm ball satisfying $\|\mathbf{S}\|_2\leq 1$. 
The projection is performed by singular-value decomposition of $\mathbf{S}$ and restricting singular values that exceed unity.

The T-matrix models were optimized for approximately $10^2$ epochs, using batch sizes of order $10^2$ and learning rates in the range $10^{-4}$ to $10^{-3}$.

\paragraph*{Direct optimization of the pixelated particle.}
For the physically realizable particle, the optimization is performed directly in the space of material distributions rather than by first optimizing a T-matrix and subsequently fitting a particle to it.
The trainable variables are the relative-permittivity values $\varepsilon(r,\theta)$ of the pixels forming the axially symmetric particle.
For every forward evaluation, the T-matrix corresponding to the current permittivity distribution is calculated using the IITMM described in Sec.~\ref{sec:iitmm}.
This T-matrix is then used in exactly the same optical classification pipeline as above to obtain the detector signals and the cross-entropy loss.
Since the T-matrix calculation is differentiable with respect to the material parameters, gradients of the classification loss are propagated through the IITMM calculation directly to the permittivity pixels.
For the non-absorbing particles considered here, $\operatorname{Im}\varepsilon=0$ and 
$\operatorname{Re}\varepsilon$ is constrained to the allowed range $[1;+\infty)$.

The direct particle optimization is computationally more demanding because the particle T-matrix is recomputed during every forward pass. Accordingly, these optimizations were performed for order of $10^2$ epochs using batches of order $10^3$ samples.

\paragraph*{Artificial neural network.}
For ANN  training optimization, we use the Adam optimizer~\cite{kingma2015adam} with a learning rate of order $10^{-4}$ and a batch size of $256$ for 50 epochs. Similar splitting was used for training as for the T-matrix and pixel particle optimization.




